\documentclass{sig-alternate-05-2015}
\usepackage{graphicx}
\usepackage{subfig}
\usepackage[hyphens]{url}
\begin{document}






%

\title{Visual Themes and Sentiment on Social Networks To Aid First Responders During Crisis Events}

\numberofauthors{1} 
\author{
\alignauthor
Prateek Dewan\textsuperscript{1}, Varun Bharadhwaj\textsuperscript{2}\thanks{These authors contributed equally to the paper.}, Aditi Mithal\textsuperscript{1}$^*$, Anshuman Suri\textsuperscript{1}$^*$,\\Ponnurangam Kumaraguru\textsuperscript{1}\\
       \affaddr{\textsuperscript{1}Indraprastha Institute of Information Technology - Delhi (IIIT-D)}\\
       \affaddr{\textsuperscript{2}National Institute of Technology, Tiruchirappalli, India}\\
       \email{\textsuperscript{1}\{prateekd,aditim13122,anshuman14021,pk\}@iiitd.ac.in, \textsuperscript{2}var6595@gmail.com}
}
\date{30 July 1999}

\maketitle
\begin{abstract}

Online Social Networks explode with activity whenever a crisis event takes place. Most content generated as part of this activity is a mixture of text and images, and is particularly useful for first responders to identify popular topics of interest and gauge the pulse and sentiment of citizens. While multiple researchers have used text to identify, analyze and measure themes and public sentiment during such events, little work has explored visual themes floating on networks in the form of images, and the sentiment inspired by them. Given the potential of visual content for influencing users' thoughts and emotions, we perform a large scale analysis to compare popular themes and sentiment across images and textual content posted on Facebook during the terror attacks that took place in Paris in 2015. 
Using state-of-the-art image summarization techniques, we discovered multiple visual themes which were popular in images, but were not identifiable through text. We uncovered instances of misinformation and false flag (conspiracy) theories among popular image themes, which were not prominent in user generated textual content, and can be of particular interest to first responders. 
Our analysis also revealed that while textual content posted after the attacks reflected negative sentiment, images inspired positive sentiment. 
To the best of our knowledge, this is the first large scale study of images posted on social networks during a crisis event.

\end{abstract}

%
%


%
%

%
%
\printccsdesc



\section{Introduction}

\begin{center}
\emph{``Use a picture. It's worth a thousand words.''}
\begin{flushright}
\emph{Tess Flanders, 1911}
\end{flushright}
\end{center}

The last decade has witnessed a revolution in communication technology with the introduction of Online Social Networks like Facebook and Twitter. Especially during crisis events like earthquakes, bomb blasts, terror attacks etc., people switch to social media to share updates, experiences, and stay up to date~\cite{fraustino2012social,palen2008online}. All of this user generated content makes OSNs a crucial source of information for first responders to get a sense of the pulse and sentiment of the public on a global scale in almost real time.

\begin{figure}[!h]
\begin{center}
\fbox{\includegraphics[scale=0.4]{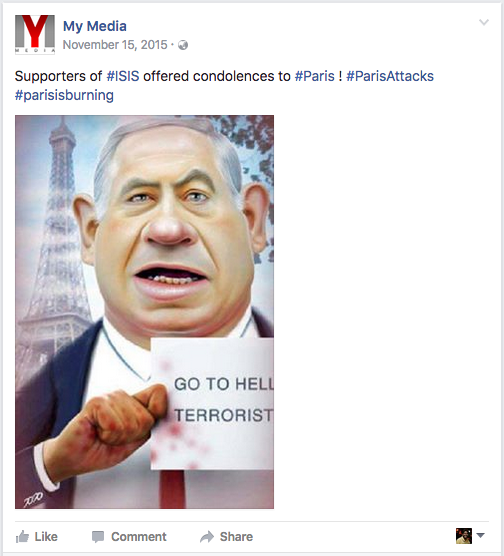}}
\end{center}
\vspace{-10pt}
\caption{
Example of a Facebook post where sentiment associated with the post text is in contrast with the sentiment associated with the text embedded in the image.
}
\label{fig:post_example}
\vspace{-10pt}
\end{figure}

In the last few years, multiple researchers have analyzed the textual part of the content to identify and analyze sentiment and popular topics of discussion among the masses on OSNs during crisis event, for example,~\cite{acar2011twitter,gupta2012identifying,hughes2009twitter,rudra2016summarizing,thelwall2011sentiment}. Sentiment and topic analysis of textual content on OSNs is widely used by researchers and first responders to gauge the pulse of citizens and make decisions accordingly. However, all OSN content is not necessarily in textual format. Researchers have reported high percentage of posts generated during real world events to contain only images, and no text at all.~\footnote{\url{http://precog.iiitd.edu.in/blog/2016/08/imagesononlinesocialmedia/}} This points to the fact that the methodology adopted for most of the aforementioned research misses out on such sections of content, which do not contain text. Moreover, even if a post contains text along with an image, the text may not be representative of the topic or sentiment depicted by the accompanying image. Consider the post in Figure~\ref{fig:post_example} for instance. While sentiment analysis on text would reveal positive sentiment for this post, the sentiment associated with the image is contrasting. Considering the human brain's affinity towards visual content,~\footnote{\url{http://changingminds.org/explanations/learning/active_learning.htm}} it is likely that the pulse and sentiment of user generated content as perceived by researchers through text, differs from the true sentiment, since most past research does not consider images to draw inferences. 

In this paper, we study a large dataset of over 57,000 images posted on Facebook during the terrorist attack in Paris in November, 2015. We employ state-of-the-art image analysis techniques for mining and measuring the themes and sentiment of content posted through images on a large scale, and compare the findings with similar analysis performed on textual content. Results of this measurement reveal sizeable differences in topics, themes, and sentiment drawn from images and text. We observed that textual content embedded in images, as well as text contained in posts depicted negative sentiment. On the other hand, images in general were found to inspire positive sentiment. Upon manual inspection, we observed that this contrasting behavior was largely due to the popularity of images offering support and solidarity to the victims of the attacks. Using a Convolutional Neural Network (CNN) based image summarization model, we extracted visual themes from images and found that 
two of the top 10 most popular themes among images 
were associated with instances of misinformation and were not prominent in textual content. 
Further, textual content extracted from images revealed multiple (potentially sensitive) topics associated with ``refugees'', ``passports'', etc. which were popular in image text, but not in post text. 

These findings indicate the presence of useful information in the form of images posted on OSNs during crisis events, which haven't been widely explored in literature. Such information can be vital for first responders to get a better sense of the pulse and sentiment of the masses, and identify and monitor instances of misinformation and sensitive information flowing on the network in the form of images. 
To the best of our knowledge, this is the first large scale study to understand visual themes and sentiment on social networks during crisis events. Further, the resulting methodology developed during our analysis scales to a generalizable model that can be applied to any similar crisis event in the future. 

{\bf About the event: }
A series of coordinated terrorist attacks took place in Paris on November 13, 2015 at 21:20 Central European Time. Suicide bombers and gunmen attacked a stadium in Saint-Denis, Paris. This was followed by mass shootings, and a suicide bombing, at cafes and restaurants. Gunmen carried out another mass shooting and took hostages at a concert in the Bataclan theatre, leading to a stand-off with police. The attackers were shot or blew themselves up when police raided the theatre. A total of 130 people were killed, and 368 others were injured. News about the event spread instantly on all OSN platforms including Facebook. Hundreds of users posted live pictures of the event, and thousands posted messages offering condolence and support.~\footnote{\url{http://www.bbc.com/news/world-europe-34818994}} 

\section{Related Work}

There exists prior literature in the space of studying images during crisis events on a small scale on OSNs, as well as studying crisis events on OSNs otherwise. Our work contributes towards enhancing the work done in both these categories, as we discuss below.

\subsection{Images on OSNs during crisis events} \label{sec:relatedwork_images}

Multiple ~researchers have attempted to study images posted on OSNs to analyze crisis events. Gupta et al. attempted to identify and characterize the spread of fake images on Twitter during Hurricane Sandy in 2013. Although the paper was focused on identifying fake images, the methodology adopted by the authors did not involve image analysis. Authors manually identified a set of fake images from news articles and blogs, and used the URLs of these fake images to expand their dataset of tweets containing fake images. This dataset was used to extract user and tweet level features to automatically identify tweets containing fake image URLs from tweets containing real image URLs~\cite{gupta2013faking}. Vis et al. conducted an exploratory analysis of images shared on Twitter during the 2011 UK riots. Similar to Gupta et al.'s approach, authors manually classified images into 14 categories for characterization~\cite{vis2013twitpic}. More similar work includes empirical analysis of Twitter images during the 2012 Israeli-Hamas conflict, where authors examined images shared by two Twitter accounts over a 2 month time frame. A total of 243 images were captured, and studied manually to discover prominent themes and frames, human characters, etc. present in the images~\cite{seo2014visual}. Kharroub et al. studied 581 Twitter images from the 2011 Egyptian revolution and found more images depicting crowds and protest activity as compared to images depicting violent content. In addition to most prominent visual themes, authors of this work tried to find whether user information helps in predicting image retweets, and whether image themes vary across different phases of the event~\cite{kharroub2015social}.

As evident from prior research, images play a crucial role in measuring public sentiment during crisis and mass emergency events like terror attacks, and in cases of detecting online radicalization. All aforementioned research however, is restricted to small scale, because of the manual effort involved in measurement and analysis. The use of images for analyzing events on a large scale remains largely unexplored. We attempt to bridge this gap to an extent, by exploring automated methods to extract meaningful information from images posted on Facebook during the Paris Attacks in 2015.

\subsection{Crisis event related studies on OSNs}

Numerous researchers have looked at textual content to study crisis events on OSNs. 
Hughes et al. studied the use of the Twitter social network during four emergency events, and compared how this behavior was different from general Twitter use. Authors found that Twitter messages sent during these types of events contained more displays of information broadcasting and brokerage as compared to general Twitter messages. Textual features like \emph{replies}, URLs, and presence of certain keywords were used to draw these findings~\cite{hughes2009twitter}. Gupta et al. presented a study to identify and characterize communities from a set of users who post messages on Twitter during three major crisis events in that took place in 2011. Authors used textual content similarity in addition to link (network) and location similarity to identify clusters of users similar users~\cite{gupta2012identifying}. Rudra et al. proposed a novel framework to assign tweets posted during mass emergency events into different situational classes, and then summarize those tweets. Similar to Hughes et al's approach of using textual features, authors of this work also extracted features like numerals, nouns, locations, verbs, etc. present in tweet text to identify and extract event summary~\cite{rudra2016summarizing}. Thelwall et al. studied sentiment of English tweets during a month long period and found that popular events were normally associated with increases in negative sentiment strength. Authors completely relied on tweet text to extract sentiment strength and draw inferences~\cite{thelwall2011sentiment}.


All aforementioned research used textual content to study events on OSNs and draw inferences, thereby missing out on a large section of content pertaining to images. As discussed previously, researchers have tried to look at images on OSNs using manual techniques, and reported interesting findings. The aforementioned research highlights the need for automated large scale techniques to study and mine images to extract sentiment, themes, and other similar useful information that can be used by researchers to better understand the users' reactions with respect to crisis events on OSNs and draw more accurate inferences.


\section{Methodology} \label{sec:methodology}

\subsection{Data collection}

We collected data using Facebook's Graph API Search endpoint~\cite{Facebook-Developers:2013} between November 14, and November 25, 2015. Until April 2015, Facebook provided developers with the capability to search for public posts given a keyword using the Graph API version 1.0. This feature was disabled with effect from April 30, 2015, with the introduction of version 2.0 of the Graph API. However, in early 2015, a vulnerability discovered in the Graph API highlighted that access tokens generated by Facebook's mobile apps had some extra permissions which were not documented by Facebook.~\footnote{\url{http://www.7xter.com/2015/02/how-i-hacked-your-facebook-photos.html}} On further exploration, we discovered that access tokens generated by Facebook's mobile apps were able to access the post search endpoint of the Graph API version 1.0 even after April 30, 2015. We used this work-around to query and collect data from the Graph API in November 2015. Table~\ref{tab:data_stats} provides the detailed description of our dataset.

\begin{table}[!h]
\vspace{-5pt}
\small
    \begin{tabular}{l|l}
    \hline
    Keywords used               & \#ParisAttacks, \#PrayForParis \\
    Unique posts                & 131,548                        \\
    Unique users                & 106,275                        \\
    Posts with images           & 75,277                         \\
    Unique users posting images & 67,570                         \\
    Total images extracted	& 57,748			\\
    Unique images extracted     & 15,123                         \\ \hline
    \end{tabular}
\vspace{-10pt}
\caption{Descriptive statistics of our dataset we collected from Facebook during the Paris Attacks in 2015.}
\vspace{-5pt}
\label{tab:data_stats}
\end{table}

The Graph API returns posts in JSON format (JavaScript Object Notation), and each post has a \emph{type} associated with it. We filtered out all posts of \emph{type ``photo''} and re-queried the Graph API in February 2016 to obtain the actual images in these posts. Upon re-querying the Graph API, we noticed that some of the posts had been deleted from Facebook, and were no longer accessible. Also, posts with entire photo albums were also categorized as \emph{type ``photo''}, and images inside these albums were not directly accessible via the API. Eventually, we were able to collect 57,748 images in total. We identified duplicates using the difference hash (dHash) image hashing algorithm~\footnote{\url{http://www.hackerfactor.com/blog/?/archives/529-Kind-of-Like-That.html}} and obtained a total of 15,123 unique images (Table~\ref{tab:data_stats}).

\subsection{Image characterization} \label{sec:image_label}

Understanding and interpreting images is a complex task. 
As discussed in Section~\ref{sec:relatedwork_images}, past research on studying the role of images on OSNs has largely relied on manual methods to perform measurement studies on images~\cite{gupta2013faking,redondo2016image,vis2013twitpic}. This methodology is time consuming and not scalable for bigger datasets containing more than a few hundred images. With millions of images generated on OSNs every day~\footnote{\url{http://www.businessinsider.in/Facebook-Users-Are-Uploading-350-Million-New-Photos-Each-Day/articleshow/22709734.cms}}, manually looking at images is a futile way to understand visual content and draw any meaningful conclusions. 

To overcome this drawback, we attempt to use automated methods to characterize images in our dataset. We use TensorFlow implementation of Google's Inception-v3 model for image classification~\cite{tensorflow-inceptionv3:2015}. Inception-v3 is trained for the ImageNet Large Visual Recognition Challenge using the data from 2012, and tries to classify images into 1,000 classes~\cite{ILSVRC15}. This model reports a state-of-the-art top-1 error rate of 17.2\%, signifying that the most probable label predicted by this model is correct in 82.8\% of the cases in the test dataset. However, it is important to note that the generated label comes from a fixed set of 1,000 labels, which is not large enough for characterizing the wide variety of images we usually come across on social networks in practice. To establish the accuracy of this model on our dataset, we recruited human annotators and got a random sample of 2,545 unique images annotated by two or more annotators. The job of each annotator was to mark whether they \emph{agreed} with the label generated by the Inception-v3 model for the given image, or not. Using majority voting, we found that the model achieved an accuracy of 38.87\% on our data sample. Given that there are 1,000 possible output labels, this accuracy value is much better than random guessing; assuming equal class sizes, the probability of a random guess being correct is 0.1\%. However, through a small manual exercise, we were able to boost this accuracy significantly.

\begin{figure}[!ht]
  \centering
  \subfloat[What a ``bolo tie'' looks like]{\includegraphics[width=0.23\textwidth]{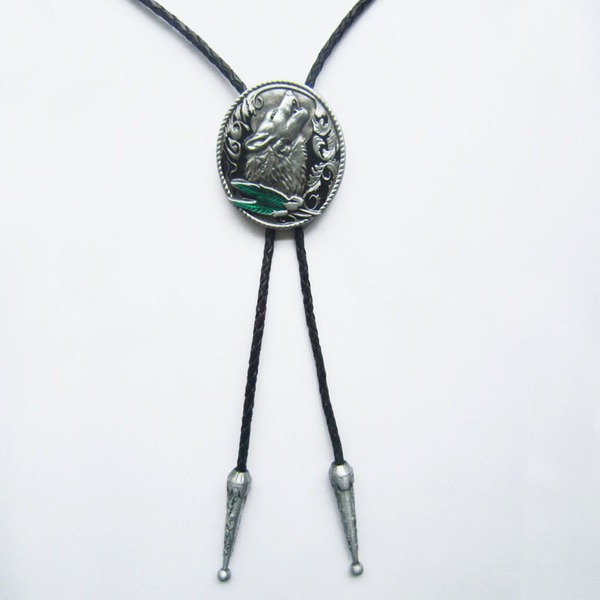}}
  \hspace{5pt}
  \subfloat[Jean Jullien's ``Peace for Paris'' symbol]{\includegraphics[width=0.23\textwidth]{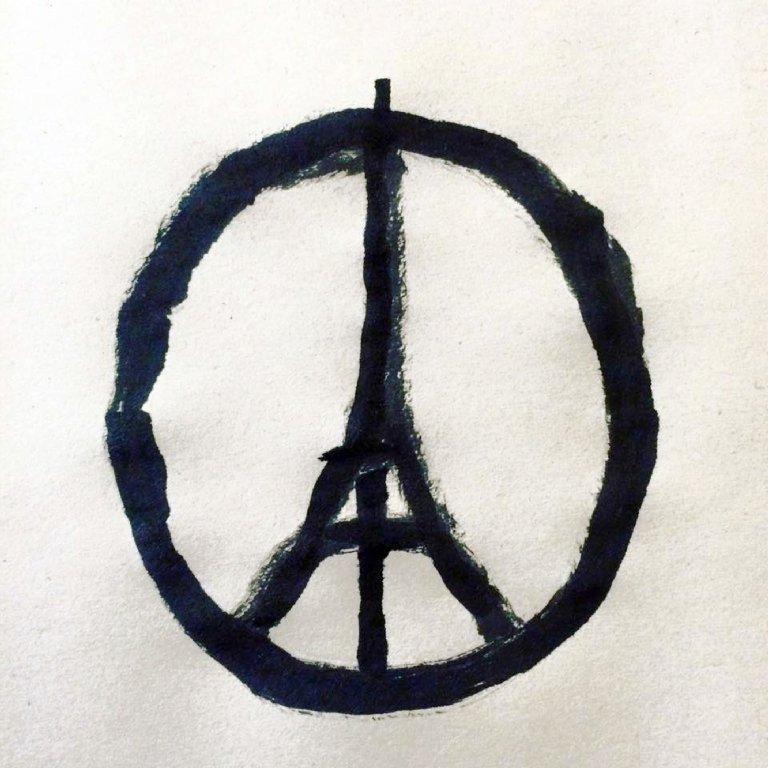}}
  \caption{Visual similarity between a bolo tie and the famous `Peace for Paris' symbol. This similarity (in addition to others) is captured by the Inception-v3 model, and can be exploited to increase accuracy.}
\label{fig:bolotie}
\end{figure}

We hypothesized that images in our dataset with the same labels are highly likely to be similar, regardless of the labels associated with them being correct or not. This is because of the fact that CNNs model the vision system in animals, and are likely to group similar images together. We tested this hypothesis for the images associated with the top 20 most frequently occurring labels in our dataset, and found it to hold true. For example, the ``Peace for Paris'' symbol created by French graphic designer Jean Jullien, was labeled ``bolo tie'' by the model (Figure~\ref{fig:bolotie}). Using this observation, we renamed 8 out of the top 20 labels to better suit the images under each of these labels. Table~\ref{tab:tag-tweaks} shows the 8 out of the 20 most common labels that were generated by the Inception-v3 model for our dataset, and the labels we replaced them with. This exercise of renaming labels boosted the accuracy of the model to 51.34\% on our random sample. We used these modified labels for our analysis.

\begin{table*}[!ht]
    \begin{tabular}{l|l|p{4cm}}
    \hline
    Label generated by Inception-v3                    & Label they were replaced with             & No. of occurrences in our dataset of unique images \\ \hline
    book jacket, dust cover, dust jacket, dust wrapper & Poster                                    & 1,024                             \\
    bolo tie, bolo, bola tie, bola                     & Jean Jullien's ``Peace for Paris'' symbol & 350                               \\
    church, church building                            & Eiffel tower                              & 258                               \\
    obelisk                                            & Eiffel tower                              & 210                               \\
    envelope                                           & Poster                                    & 204                               \\
    stupa, tope                                        & Eiffel tower                              & 173                               \\
    drilling platform, offshore rig                    & Eiffel tower                              & 137                               \\
    radio telescope, radio reflector                   & Eiffel tower                              & 68                                \\ \hline
    \end{tabular}
\vspace{-5pt}
\caption{List of labels generated by the Inception-v3 model, and the labels they are renamed with. This renaming process boosted the accuracy of the model on our dataset by almost 13\%.}
\label{tab:tag-tweaks}
\vspace{-10pt}
\end{table*}

\subsection{Text extraction from images} \label{sec:ocr}

\begin{figure}[!h]
\begin{center}
\fbox{\includegraphics[scale=0.35]{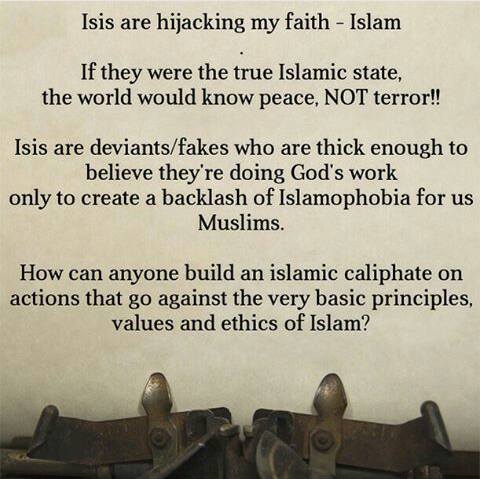}}
\end{center}
\vspace{-10pt}
\caption{Example of text embedded in an image posted on Facebook during the Paris Attacks in 2015. We found thousands of images containing text in our dataset.}
\label{fig:text_in_image}
\vspace{-10pt}
\end{figure}

Past studies on analysing topics, events, sentiment, etc. on OSNs have been largely limited to using textual content generated by users to draw inferences~\cite{gupta2012identifying,hughes2009twitter,rudra2016summarizing,thelwall2011sentiment}. This technique however, misses out on a large section of textual information embedded in images, making it prone to missing out on being able to capture the complete picture. Figure~\ref{fig:text_in_image} shows an example of textual content embedded in an image in our dataset. 

We tested and evaluated two optical character recognition (OCR) libraries -- PyTesseract~\footnote{\url{https://pypi.python.org/pypi/pytesseract/}} (Python wrapper for Tesseract OCR~\footnote{\url{https://github.com/tesseract-ocr/tesseract}}), and OCRopy~\footnote{\url{https://pypi.python.org/pypi/ocropy}} -- to extract text from the images in our dataset. To compare the performance of the two libraries, we first established a ground truth dataset by manually extracting text from a random sample of 1,000 images from our dataset. We then used five string similarity metrics (Jaro Winkler distance, Jaccard index, Cosine similarity, Hamming distance, and Levenshtein distance) to compare the results produced by PyTesseract and OCRopy with the ground truth, separately. Scores from all five metrics indicated that PyTesseract performed better on our dataset than OCRopy (p-value $<$ 0.01 for all five metrics), so we used PyTesseract for our final analysis. In all, we were able to extract text from a total of 31,869 images in our dataset.

\subsection{Identifying image sentiment} \label{sec:image_sentiment}

Sentiment derived from textual content generated by users on OSNs has been widely used by researchers in various contexts~\cite{bollen2011twitter,stieglitz2012political,thelwall2011sentiment,tumasjan2010predicting}. However, few attempts have been made to understand the sentiment associated with images posted on OSNs~\cite{xu2014visual,you2015robust,yuan2013sentribute}. Studies suggests that the human brain is hardwired to recognize and make sense of visual information more efficiently.~\footnote{\url{https://www.eyeqinsights.com/power-visual-content-images-vs-text/}} Thus, it is likely that sentiment extracted from textual content alone may not be representative of the overall sentiment associated with a theme or event. To this end, we attempt to extract sentiment from images using state-of-the-art supervised learning techniques and datasets. 


The Inception-v3 model (discussed in Section~\ref{sec:image_label}) can be retrained to perform other visual recognition tasks using features extracted by the model during the training phase. This concept is known as transfer learning, and is available in the form of an open-source implementation, called Deep Convolutional Activation Feature (DeCAF).~\footnote{\url{https://www.tensorflow.org/versions/r0.8/how_tos/image_retraining/index.html}} DeCAF is a state-of-the-art deep CNN architecture for transfer learning based on a supervised pre-training phase~\cite{donahue2014decaf}. We use this open-source implementation to retrain the Inception-v3 model on the SentiBank dataset to identify image sentiment. The SentiBank database comprises of a total of half million Flickr images, extracted by querying the network using Adjective-Noun Pairs (ANPs)~\cite{borth2013sentibank}. Since noun queries such as ``dog'', ``baby'', or ``house'' do not portray a well-defined emotion, these queries were prefixed with adjectives to form ANPs like ``happy dog'', ``adorable baby'', ``abandoned house'' etc., which associate these nouns with a strong emotion. We manually segregated these ANPs (and therefore, the images associated with them) into \emph{positive} and \emph{negative} classes for binary sentiment classification, and skipped the ANPs which did not fit clearly into a \emph{positive} or \emph{negative} sentiment. This exercise left us with a total of 305,100 positive sentiment images and 133,108 negative sentiment images. We performed a 10-fold random subsampling to balance the classes, and obtain an unbiased model. For each fold, we split the dataset into three parts in a 80:10:10 ratio for training, validation, and testing respectively, and achieved a maximum accuracy of 69.8\%.

\section{Analysis and Results}

Themes and sentiment are two of the most widely studied aspects of OSN content during crisis events in literature~\cite{cheong2011microblogging,mendoza2010twitter}. We therefore focus on these two aspects of the images in our dataset, and present our findings.

\subsection{Most prominent visual themes featured misinformative images}

Using our methodology for characterizing images as described in Section~\ref{sec:image_label}, we assigned a label to all 57 thousand images we collected (Table~\ref{tab:data_stats}). Table~\ref{tab:image_labels} shows a list of the most commonly occurring image labels, along with their description according to our dataset. We manually browsed through images corresponding to each of the top 20 labels, and found that the most common types of images comprised of posters, banners, screenshots of Facebook posts, Twitter tweets, etc. Cartoons and animated posters resembling a comic book were also very popular. 
More examples include the Pray for Paris peace symbol by French artist Jean Jullien (label: Bola Tie), images of candles and lamps offering support to the victims of the attacks (label: Candle waxlight), and images of the Eiffel Tower under varying lights, angles, and from various distances, that became very popular (labels: Obelisk, Crane, etc.).

\begin{table}
\small
    \begin{tabular}{p{2.1cm}|p{0.7cm}|p{4.5cm}}
    \hline
    Label                    & Count	& Description                                                          \\ \hline
    Website                  & 12,416	& Images of posts, tweets, banners, etc.                        \\
    Book jacket              & 5,383	& Posters, banners, etc.                                               \\
    Comic book               & 3,803	& Cartoons, animated posters                                           \\
    Fountain                 & 1,264	& Fountains at various locations                                       \\
    Envelope                 & 1,248	& Posters, banners, etc.                                               \\
    Suit (clothing)          & 1,246	& People wearing suit-like clothes                              \\
    Stage                    & 1,135	& Stages during public speeches, mass gathering events, etc.           \\
    Candle waxlight  	     & 1,021	& Lit candles and lamps offering support to victims                    \\
    Malinois                 & 995	& Police dog who died during the attack                                \\
    Scoreboard               & 971	& Images of sports stadium                                             \\
    Microphone               & 906	& Individuals addressing the masses, reporters, etc. using microphones \\
    Menu                     & 868	& Images containing well formatted text                                \\
    Bola Tie                 & 781	& Peace for Paris symbol originally created by Jean Jullien            \\
    Bell cot                 & 745	& Various buildings                                     \\
    Jersey, T-shirt          & 743	& People wearing t-shirts                                              \\
    Crane                    & 677	& Images of Eiffel Tower during twilight                               \\
    Memorial Tablet          & 633	& Variety of posters, hand written messages on boards, etc.            \\
    Church                   & 629	& Dark and grey scale images of Eiffel Tower                           \\
    Palace                   & 586	& Large buildings, including Eiffel Tower from a distance              \\
    Obelisk                  & 547	& Eiffel Tower                                                         \\ \hline
    \end{tabular}
\caption{Top 20 most common image labels in our dataset. Majority of the images comprised of posters, banners, art work, etc. offering support for victims.}
\label{tab:image_labels}
\vspace{-15pt}
\end{table}

\begin{figure}[!ht]
  \centering
  \subfloat[Diesel the dog who was allegedly killed by terrorists]{\includegraphics[width=0.3\textwidth]{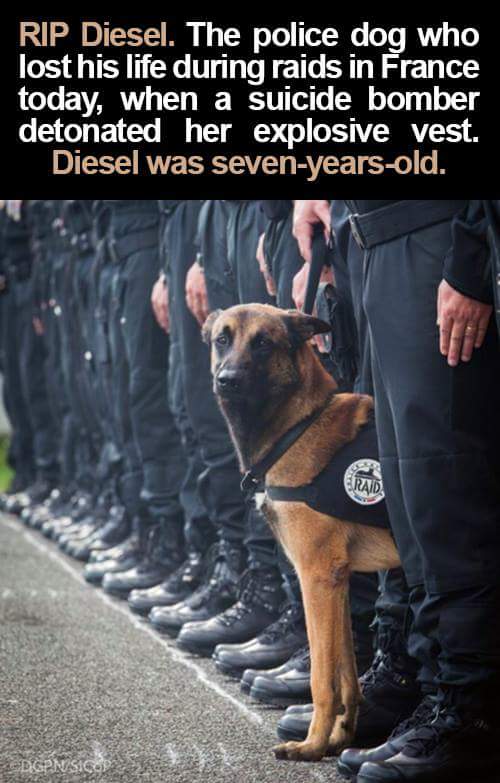}\label{fig:dieselrumor}}
  \hspace{5pt}
  \subfloat[Zouheir, the security guard who was claimed to have stopped a terrorist from entering the stadium]{\includegraphics[width=0.45\textwidth]{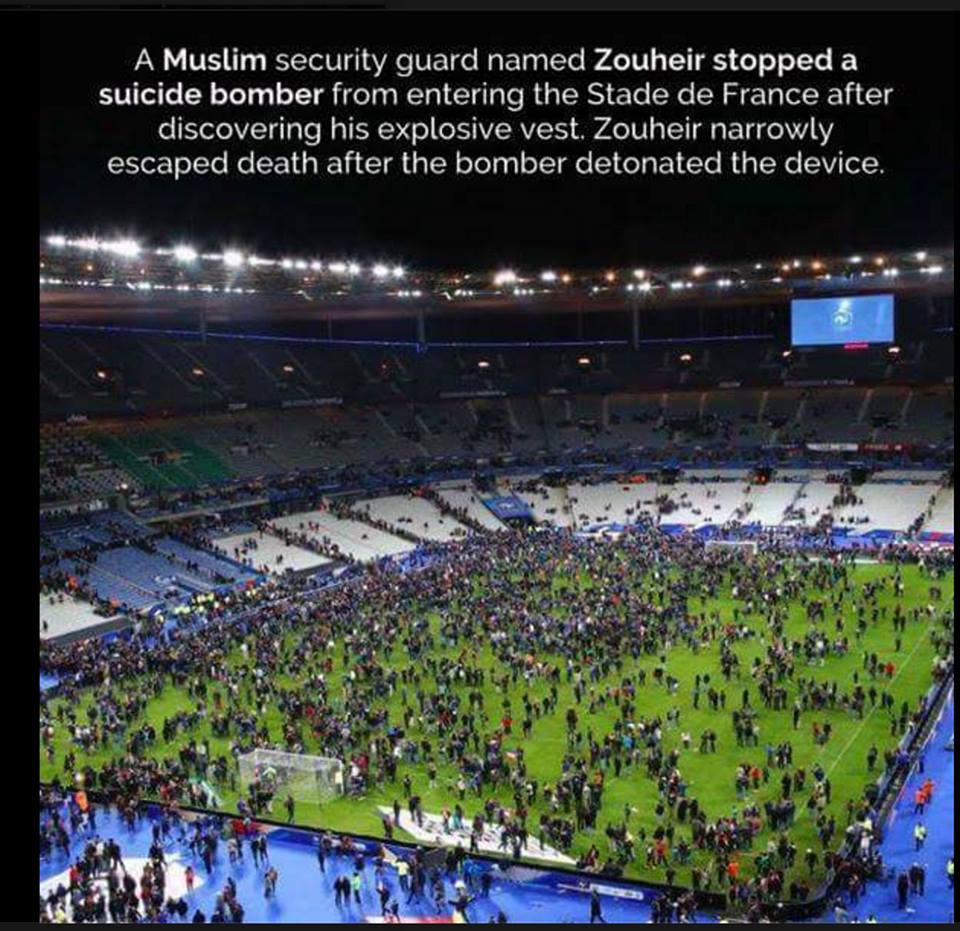}\label{fig:zouheir}}
  \caption{
Rumors spread on Facebook in the form of images during the Paris Attacks in 2015. We used CNN based image summarization techniques to identify image themes and discovered that some of the most popular image themes were associated with rumors.
}
\label{fig:rumors}
\vspace{-10pt}
\end{figure}

We were also able to identify some peculiar topics and themes which were popular on the network during the event. The ``Malinois'' label appearing in the top 20 (see Table~\ref{tab:image_labels}) corresponded to the breed of the police dog that died during the attacks, and evidently became very popular. However, the cause of death of the dog was incorrectly quoted in multiple such images. 
Figure~\ref{fig:dieselrumor} shows one such picture of the police dog and states that the dog was killed when a suicide bomber detonated her explosive vest. However the real cause of the dog's death, as later clarified by French police, was multiple gunshot wounds caused by the French police forces' ``Brenneke'' bullets.~\footnote{\url{http://www.dailymail.co.uk/news/article-3446511/Confirmed-Diesel-hero-police-dog-Paris-attacks-shot-dead-wounded-innocent-neighbours-reckless-shooting.html}} We collected all such images quoting misinformation in our dataset and found that these images had gathered over 1.1 million likes, 321 thousand shares, and 38 thousand comments.

Similarly, one of the blasts during the attacks took place outside a football stadium, whose pictures quoted incorrect information and became viral. These pictures were captured using the ``Scoreboard'' label; manual verification of images marked with this label revealed that most of these images captured the sports stadium. Figure~\ref{fig:zouheir} shows one such picture of the stadium and states that a Muslim security guard named Zouhier stopped a suicide bomber from entering the Stade de France stadium, thus saving hundreds of innocent lives of people inside the stadium. It was later confirmed by BBC that it wasn't actually him who turned away the bomber. Instead, Zouheir was stationed elsewhere in the stadium, and related what he heard from colleagues who were closer to the bomb blast.~\footnote{\url{http://www.bbc.com/news/blogs-trending-34845882}} All instances of this misinformative image in our dataset garnered over 21 thousand likes, 11 thousand shares, and 450 comments.

This technique of automatic identification of themes and topics from images on a large scale can be particularly helpful for first responders to identify popular instances of misinformation spread through images. Slight modifications to Convolutional Neural Network (CNN) based labeling models like Inception-v3, can aid in identifying potentially harmful and sensitive content such as guns, blood, etc. in images, and help monitor the flow of such images, and react in a timely manner, if needed.

\subsection{Text embedded in images featured sensitive topics and reflected negative sentiment}

Applying optical character recognition (OCR) on images in our dataset revealed 31,869 images (55\% of all images in our dataset) which contained text embedded in them (as described in Section~\ref{sec:ocr}). 

{\bf Prominent topics:} Table~\ref{tab:top_topics} shows a mutually exclusive set of the 20 most frequently occurring relevant words in the text we extracted from images and posts. We picked 500 most commonly occurring words in images that were not present in post text, and vice versa, to identify prominent themes among image and post text independently. We noticed that the most commonly occurring words among image and post text had less than 45\% overlap, highlighting that popular words among image text were considerably different from those in post text. 

Text extracted from posts was dominated by words like \emph{``prayers'', ``prayfortheworld'', ``life'', ``support'', ``god''} etc., depicting support and solidarity for the victims. Text extracted from images however, revealed some potentially sensitive topics like \emph{refugees, passports}, etc. which were not amongst the most talked about topics in post text. The terms \emph{``refugees''} and \emph{``texas''} in Table~\ref{tab:top_topics} captures the scanned image of a letter from the Governor of Texas to the president of the United States, which went viral. The letter stated that Texas would not accept any refugees from Syria in the wake of the attacks. Other similar examples of images containing text comprised of quotes from famous personalities including The Dalai Lama, Stacy Washington, Barrack Obama, Martin Luther King Jr., etc. We also uncovered a popular conspiracy theory surrounding the Syrian ``passports'' that were found by French police near the bodies of terrorists who carried out the attacks, and were allegedly used to establish the identity of the attackers as Syrian citizens. Text embedded in images depicting this theme questioned how the passports could have survived the heat of the blasts and fire. 
This conspiracy theory was then used by miscreants to label the attacks as a \emph{false flag} operation, influencing citizens to question the policies and motives of their own government. The popularity of such memes on OSN platforms can have undesirable outcomes in the real world, like protests and mass unrest. It is therefore vital for first responders to be able to identify such content and counter / control its flow to avoid repercussions in the real world.

All such popular textual content in the form of images is a rich source of information which is capable of playing an important role in driving the pulse of the audience, but has been largely untapped in literature so far. Interestingly, 8,273 of these 31,869 images (25.95\%) did not contain any user generated textual content otherwise, indicating that most prior work on event analysis using text on OSNs would have entirely missed this set of data during their analysis, as discussed previously~\cite{hughes2009twitter,gupta2012identifying,thelwall2011sentiment,rudra2016summarizing}.

\begin{table}
    \begin{tabular}{l|lp{1cm}|lp{1cm}}
    \hline
    ~      & \multicolumn{2}{l|}{Top words in posts} & \multicolumn{2}{l}{Top words in images} \\ \hline
    ~ 	   & Word                & Norm. freq. & Word                & Norm. freq. \\ \hline
    1.     & retweeted           &     0.0055 & house               &     0.0045       \\
    2.     & time                &     0.0052 & safety              &     0.0044      \\
    3.     & prayers             &     0.0050 & washington          &     0.0042      \\
    4.     & news                &     0.0047 & sisters             &     0.0039      \\
    5.     & prayfortheworld     &     0.0044 & learned             &     0.0038 \\
    6.     & life                &     0.0043 & mouth               &     0.0038 \\
    7.     & let                 &     0.0042 & stacy               &     0.0037 \\
    8.     & support             &     0.0042 & passport            &     0.0037 \\
    9.     & god                 &     0.0040 & americans           &     0.0036 \\
    10.    & war                 &     0.0039 & refugee             &     0.0035 \\
    11.    & thoughts            &     0.0038 & japan               &     0.0028 \\
    12.    & need                &     0.0038 & texas               &     0.0027 \\
    13.    & last                &     0.0037 & born                &     0.0026 \\
    14.    & lives               &     0.0037 & dear                &     0.0026 \\
    15.    & said                &     0.0034 & syrians             &     0.0026 \\
    16.    & place               &     0.0034 & similar             &     0.0025 \\
    17.    & country             &     0.0033 & deadly              &     0.0025 \\
    18.    & city                &     0.0032 & services            &     0.0025 \\
    19.    & everyone            &     0.0032 & accept              &     0.0025 \\
    20.    & live                &     0.0032 & necessary           &     0.0025 \\ \hline
    \end{tabular}
\vspace{-5pt}
\caption{Mutually exclusive set of 20 most frequently occurring relevant keywords in post and image text, with their normalized frequency. We identified some potentially sensitive topics among image text, which were not present in post text. Word frequencies are normalized independently by the total sum of frequencies of the top 500 words in each class.}
\label{tab:top_topics}
\vspace{-10pt}
\end{table}

\begin{figure}[!h]
\begin{center}
\fbox{\includegraphics[scale=0.4]{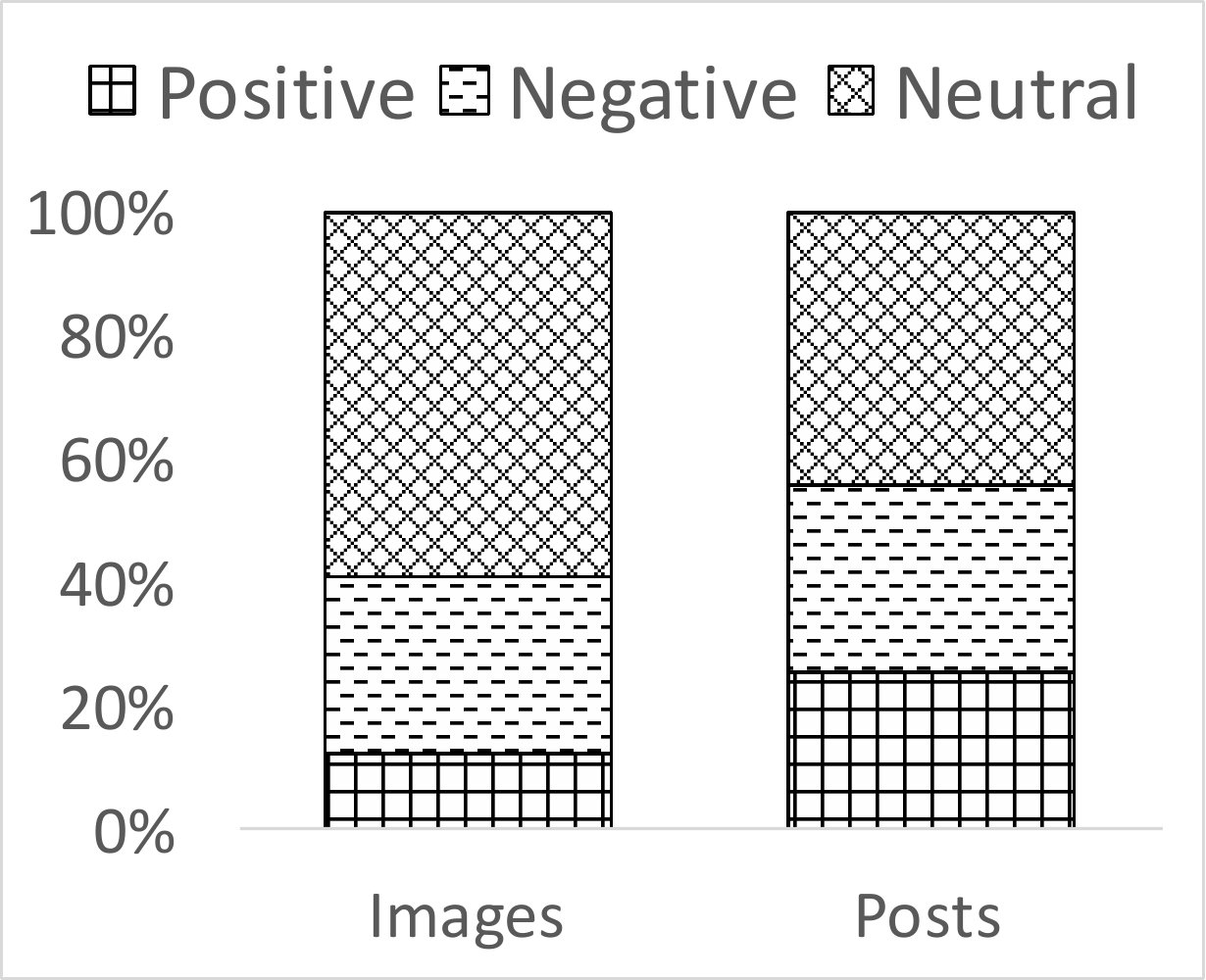}}
\end{center}
\vspace{-13pt}
\caption{Distribution of positive, negative, and neutral text emotion in our dataset of images containing text, and posts. We found more negative emotion in both categories, but the magnitude of negative emotion as compared to positive emotion was higher in images.}
\label{fig:images_post_sentiment}
\vspace{-2pt}
\end{figure}

{\bf Text sentiment:} Researchers in the past have looked at text sentiment to draw inferences about the overall sentiment and emotion of users about a topic on OSNs. However, as we just discussed, there is a large volume of textual content flowing through the network in the form of images, which differs from conventional textual content posted by users. Since most modern content monitoring techniques also focus on textual content, obfuscating sensitive textual content like hate speech and propaganda by embedding it in images is a lucrative way for malicious entities to avoid detection. Thus, we hypothesize that the sentiment of text embedded in images would be very different from the sentiment of textual content posted by users in the conventional form. To confirm our hypothesis, we employed Linguistic Enquiry and Word Count (LIWC)~\cite{pennebaker2007development} to determine and compare the emotion of the image text and post text in our dataset. We found that text embedded in images was negative on average, and twice in magnitude as post emotion (Mann Whitney U test: $p-value < 0.01$). Figure~\ref{fig:images_post_sentiment} shows the distribution of emotions across the two classes (image text and post text). Although we found more negative emotion in both images and posts as compared to positive emotion, the magnitude of negative emotion as compared to positive emotion was much higher in images as compared to text. We also noticed that positive emotion in posts was 2.6 times higher in magnitude than positive emotion in images. For negative emotion, this magnitude dropped down to 1.25.

These results indicate that textual content flowing on the network in the form of images is a critical source of information that needs to be taken into consideration and analyzed thoroughly while making a judgement on the pulse and sentiment of the audience about any topic or event.

Differences in text sentiment obtained from images and posts highlight that studying images is essential to correctly gauge the sentiment of users about a topic on OSNs. With the large volume and popularity of images on OSNs in recent times, it is imminent that results drawn from text alone fail to accurately capture the pulse of the audience. Especially during crisis events, incorrect judgement of citizen sentiment can lead to undesirable consequences in the real world, like mass unrest and public protests, which can otherwise be stopped by law and order agencies if they are able to capture a complete picture of the situation in time.


\subsection{Images inspired positive sentiment}

Inferring image sentiment through the sentiment of text embedded in it (as discussed previously), is a small part of understanding the sentiment associated with an image. Text is only a part of the overall sentiment that an image may reflect. Moreover, there may be no text present in an image at all. Researchers have acknowledged the problem of understanding image sentiment, and come up with some solutions recently~\cite{xu2014visual,you2015robust,yuan2013sentribute}. Using some of the most advanced techniques in the domain of image sentiment extraction (as described in Section~\ref{sec:image_sentiment}), we performed sentiment analysis of images in our dataset.

\begin{figure}[!h]
\begin{center}
\fbox{\includegraphics[scale=0.25]{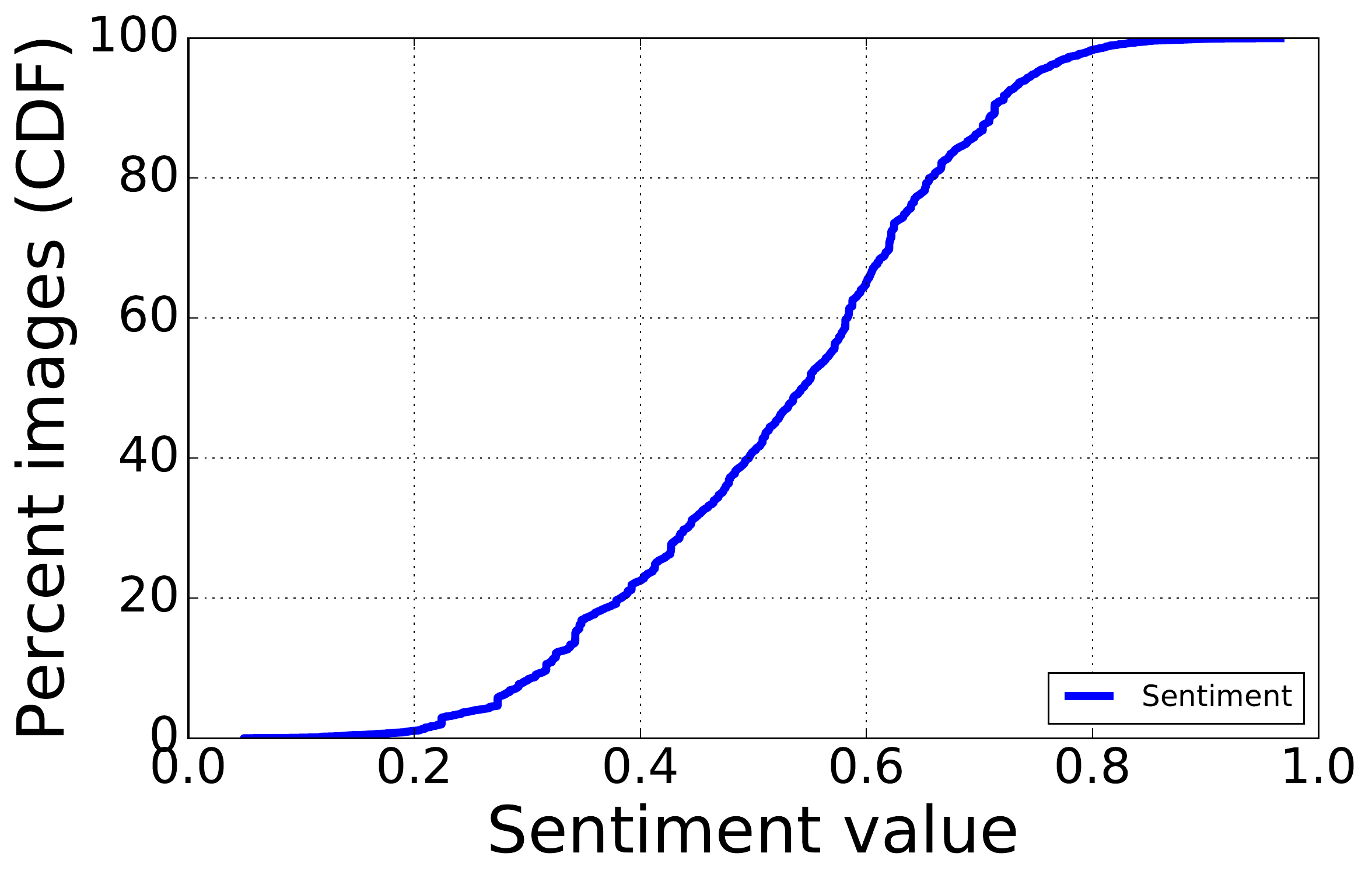}}
\end{center}
\vspace{-13pt}
\caption{
Cumulative distribution of sentiment values obtained from our image sentiment prediction model for all 57 thousand images in our dataset. A value over 0.5 depicts positive sentiment. We observed that close to 60\% of all images depicted positive sentiment.
}
\label{fig:image_senti_graph}
\vspace{-2pt}
\end{figure}

Contrary to text sentiment, we found that images, on average, portrayed a positive sentiment. Figure~\ref{fig:image_senti_graph} shows the normalized sentiment scores of all the images in our dataset. As is evident from the figure, we observed that close to 60\% of the 57 thousand images in our dataset depicted a positive sentiment. However, as discussed in Section~\ref{sec:image_sentiment}, the accuracy of our image sentiment model wasn't too high (approximately 70\%). Therefore, to verify the validity of our observations, we recruited human annotators to manually mark a small random sample of 2,545 images from our dataset as positive, negative, or neutral. Participants were also given an option to skip. Each image was annotated by at least 2 (and at most 3) participants. After removing the skipped images and using majority voting, we found that 50.95\% images were marked as positive, whereas only 16.21\% images were marked as negative. The remaining images were marked as neutral. This exercise confirmed our findings and affirmed the dominance of positive sentiment images in our dataset.

This observation can be attributed to the large number of pictures depicting support and solidarity for the victims of the attacks, which included posters, banners, people holding lit candles and lamps, the famous Peace for Paris symbol, etc. Such images inspire a positive sentiment on the viewer, as confirmed by our human annotators as well as the pre-trained sentiment prediction model.

\begin{figure}[!h]
\begin{center}
\fbox{\includegraphics[scale=0.4]{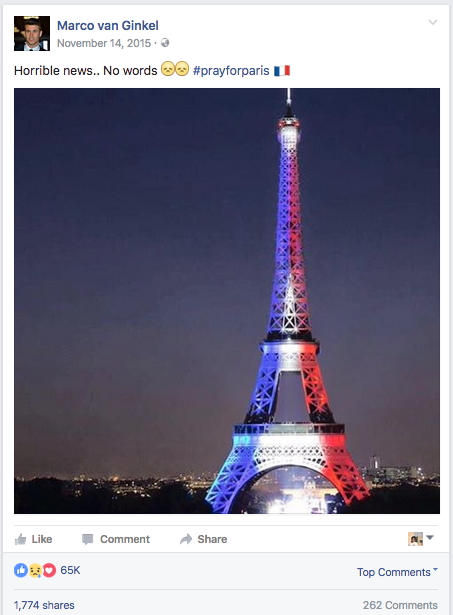}}
\end{center}
\vspace{-10pt}
\caption{
Example of a post published during the Paris attacks, showing conflicting sentiments across image and text. This post was published by a verified page, garnering over 65 thousand likes and was shared 1,774 times.
}
\label{fig:conflicting_sentiment}
\vspace{-3pt}
\end{figure}

Interestingly, we came across a substantial number of instances where image sentiment conflicted with the sentiment of the text present in the post. Consider the post shown in Figure~\ref{fig:conflicting_sentiment} for example. While the text in the post reads, \emph{``Horrible news.. No words :( :(''} reflecting highly negative sentiment, the image depicts the Eiffel tower lit up in French colors, signifying support for the victims and reflecting a positive sentiment. We observed that, out of the 19,954 posts in our dataset which contained user generated textual content as well as an image, 25.33\% of the posts (5,056 posts) had conflicting image and text sentiments. Out of these, 10.98\% of the posts (2,192 posts) contained an image depicting a negative sentiment, whereas the textual content present in the post reflected positive sentiment. Similarly, 14.35\% of the posts (2,864 posts) contained an image depicting a positive sentiment, whereas the textual content present in the post reflected negative sentiment. 

Through this analysis, we uncovered a new dimension for mining sentiment from user generated content on OSNs, which has been largely unexplored in prior OSN related literature. Our results shed light on the varying sentiment depicted by images and text during the Paris attacks. While textual sentiment analysis revealed negative sentiment, we found that images shared on Facebook during the event depicted positive sentiment. We also found a considerable proportion of posts where textual sentiment and image sentiment depicted opposite polarity. It is important to note that while text has been widely accepted in literature as a means to infer user sentiment during crisis events (and otherwise) on OSNs, the sentiment perceived by users is not restricted to text only. Instead, given the affinity of the human mind towards visual content, images are likely to contribute much more to the perceived sentiment of users as compared to text.

\section{Limitations}

We understand that our image labeling (Inception-v3) and sentiment detection models are not completely accurate. However, these models are trained using CNNs, which form the basis for one of the most advanced state-of-the-art techniques for learning features from visual media in the modern age. These models can be further improved by feeding them true positive datasets of images. Generating a big enough dataset for such models is however, a challenging task, and out of scope of this work.

Text extracted using Tesseract is limited by the performance of modern OCR techniques. We came across instances in our dataset where we were manually able to recognize text, but the OCR failed to identify this text. Most such instances involved the presence of calligraphic text, and noisy background. The percentage of such instances was low as compared to the number of images for which we were able to identify and extract text.

\section{Discussion}

Images are an integral part of OSN content, and are naturally more appealing to the human mind than text. OSN statistics explicitly support this phenomenon and makes images an interesting avenue for researchers to explore. In this paper, we discovered a new dimension of content on Facebook through images. This study highlights the vast amount of information that can be mined from images alone, by making use of some modern image analysis techniques. We also highlight how this information may be different from textual content that has been previously used in literature to infer users' pulse and sentiment. 

Pictures posted on OSNs are a critical source of information that can be useful for law and order organizations to understand public sentiment, especially during crisis events. With the enormous volume and velocity of data being generated on OSNs, it is extremely difficult to monitor visual media at present, because of lack of automated methods for measurement and analysis of such content. Through our approach, we propose a semi-automated methodology for mining knowledge from visual content and identifying popular themes and citizens' pulse during crisis events. Although this methodology has its limitations, it can be very effective for producing high level summaries and reducing the search space for organizations with respect to content that may need attention. We also described how our methodology can be used for automatically identifying (potentially sensitive) misinformation spread through images during crisis events, which may lead to major implications in the real world.

Various researchers have used text to measure the sentiment and mood of users in diverse contexts like natural calamities, politics, sports, etc. We believe that the results drawn from all these studies can be improved by taking visual content into consideration.

Brands and organizations invest heavily in social media marketing and rely on textual responses generated by users to gauge their reactions, and in turn, the performance of their products. Being able to understand the users' pulse through images is likely to help such organizations measure the response to their products much better, and cover a larger section of the audience. Moreover, while analyzing sentiment and emotion through text is largely limited by language, such a barrier does not exist for images.

\bibliographystyle{abbrv}
\bibliography{/Users/prateekdewan/Dropbox/Super_BibTex_Collection/Prateek} 

\begin{thebibliography}{10}

\bibitem{acar2011twitter}
A.~Acar and Y.~Muraki.
\newblock Twitter for crisis communication: lessons learned from japan's
  tsunami disaster.
\newblock {\em International Journal of Web Based Communities}, 7(3):392--402,
  2011.

\bibitem{bollen2011twitter}
J.~Bollen, H.~Mao, and X.~Zeng.
\newblock Twitter mood predicts the stock market.
\newblock {\em Journal of Computational Science}, 2(1):1--8, 2011.

\bibitem{borth2013sentibank}
D.~Borth, T.~Chen, R.~Ji, and S.-F. Chang.
\newblock Sentibank: large-scale ontology and classifiers for detecting
  sentiment and emotions in visual content.
\newblock In {\em Proceedings of the 21st ACM international conference on
  Multimedia}, pages 459--460. ACM, 2013.

\bibitem{cheong2011microblogging}
M.~Cheong and V.~C. Lee.
\newblock A microblogging-based approach to terrorism informatics: Exploration
  and chronicling civilian sentiment and response to terrorism events via
  twitter.
\newblock {\em Information Systems Frontiers}, 13(1):45--59, 2011.

\bibitem{donahue2014decaf}
J.~Donahue, Y.~Jia, O.~Vinyals, J.~Hoffman, N.~Zhang, E.~Tzeng, and T.~Darrell.
\newblock Decaf: A deep convolutional activation feature for generic visual
  recognition.
\newblock In {\em ICML}, pages 647--655, 2014.

\bibitem{Facebook-Developers:2013}
{Facebook Developers}.
\newblock Facebook graph api search.
\newblock {\em
  \url{https://developers.facebook.com/docs/graph-api/using-graph-api/v1.0#search}},
  2013.

\bibitem{fraustino2012social}
J.~D. Fraustino, B.~Liu, and Y.~Jin.
\newblock Social media use during disasters: a review of the knowledge base and
  gaps.
\newblock 2012.

\bibitem{gupta2012identifying}
A.~Gupta, A.~Joshi, and P.~Kumaraguru.
\newblock Identifying and characterizing user communities on twitter during
  crisis events.
\newblock In {\em Proceedings of the 2012 Workshop on Data-driven User
  Behavioral Modelling and Mining from Social Media}, DUBMMSM '12, pages
  23--26, New York, NY, USA, 2012. ACM.

\bibitem{gupta2013faking}
A.~Gupta, H.~Lamba, P.~Kumaraguru, and A.~Joshi.
\newblock Faking sandy: characterizing and identifying fake images on twitter
  during hurricane sandy.
\newblock In {\em Proceedings of WWW companion}, pages 729--736, 2013.

\bibitem{hughes2009twitter}
A.~L. Hughes and L.~Palen.
\newblock Twitter adoption and use in mass convergence and emergency events.
\newblock {\em International Journal of Emergency Management}, 6(3):248--260,
  2009.

\bibitem{kharroub2015social}
T.~Kharroub and O.~Bas.
\newblock Social media and protests: An examination of twitter images of the
  2011 egyptian revolution.
\newblock {\em New Media \& Society}, page 1461444815571914, 2015.

\bibitem{mendoza2010twitter}
M.~Mendoza, B.~Poblete, and C.~Castillo.
\newblock Twitter under crisis: Can we trust what we rt?
\newblock In {\em Proceedings of the first workshop on social media analytics},
  pages 71--79. ACM, 2010.

\bibitem{palen2008online}
L.~Palen.
\newblock Online social media in crisis events.
\newblock {\em Educause Quarterly}, 31(3):76--78, 2008.

\bibitem{pennebaker2007development}
J.~W. Pennebaker, C.~K. Chung, M.~Ireland, A.~Gonzales, and R.~J. Booth.
\newblock The development and psychometric properties of liwc2007, 2007.

\bibitem{redondo2016image}
R.~Q. Redondo.
\newblock The image use on twitter during the 2015 municipal election campaign
  in spain.
\newblock {\em Revista Latina de Comunicaci{\'o}n Social}, 71:85--107, 2016.

\bibitem{rudra2016summarizing}
K.~Rudra, S.~Banerjee, N.~Ganguly, P.~Goyal, M.~Imran, and P.~Mitra.
\newblock Summarizing situational tweets in crisis scenario.
\newblock In {\em Proceedings of the 27th ACM Conference on Hypertext and
  Social Media}, pages 137--147. ACM, 2016.

\bibitem{ILSVRC15}
O.~Russakovsky, J.~Deng, H.~Su, J.~Krause, S.~Satheesh, S.~Ma, Z.~Huang,
  A.~Karpathy, A.~Khosla, M.~Bernstein, A.~C. Berg, and L.~Fei-Fei.
\newblock {ImageNet Large Scale Visual Recognition Challenge}.
\newblock {\em International Journal of Computer Vision (IJCV)},
  115(3):211--252, 2015.

\bibitem{seo2014visual}
H.~Seo.
\newblock Visual propaganda in the age of social media: An empirical analysis
  of twitter images during the 2012 israeli--hamas conflict.
\newblock {\em Visual Communication Quarterly}, 21(3):150--161, 2014.

\bibitem{stieglitz2012political}
S.~Stieglitz and L.~Dang-Xuan.
\newblock Political communication and influence through microblogging--an
  empirical analysis of sentiment in twitter messages and retweet behavior.
\newblock In {\em System Science (HICSS), 2012 45th Hawaii International
  Conference on}, pages 3500--3509. IEEE, 2012.

\bibitem{tensorflow-inceptionv3:2015}
C.~Szegedy, V.~Vanhoucke, S.~Ioffe, J.~Shlens, and Z.~Wojna.
\newblock Rethinking the inception architecture for computer vision.
\newblock {\em CoRR}, abs/1512.00567, 2015.

\bibitem{thelwall2011sentiment}
M.~Thelwall, K.~Buckley, and G.~Paltoglou.
\newblock Sentiment in twitter events.
\newblock {\em Journal of the American Society for Information Science and
  Technology}, 62(2):406--418, 2011.

\bibitem{tumasjan2010predicting}
A.~Tumasjan, T.~O. Sprenger, P.~G. Sandner, and I.~M. Welpe.
\newblock Predicting elections with twitter: What 140 characters reveal about
  political sentiment.
\newblock In {\em Proceedings of the fourth international AAAI conference on
  weblogs and social media}, pages 178--185, 2010.

\bibitem{vis2013twitpic}
F.~Vis, S.~Faulkner, K.~Parry, Y.~Manyukhina, and L.~Evans.
\newblock Twitpic-ing the riots: analysing images shared on twitter during the
  2011 uk riots.
\newblock {\em Twitter and Society}, pages 385--398, 2013.

\bibitem{xu2014visual}
C.~Xu, S.~Cetintas, K.-C. Lee, and L.-J. Li.
\newblock Visual sentiment prediction with deep convolutional neural networks.
\newblock {\em arXiv preprint arXiv:1411.5731}, 2014.

\bibitem{you2015robust}
Q.~You, J.~Luo, H.~Jin, and J.~Yang.
\newblock Robust image sentiment analysis using progressively trained and
  domain transferred deep networks.
\newblock {\em arXiv preprint arXiv:1509.06041}, 2015.

\bibitem{yuan2013sentribute}
J.~Yuan, S.~Mcdonough, Q.~You, and J.~Luo.
\newblock Sentribute: image sentiment analysis from a mid-level perspective.
\newblock In {\em Proceedings of the Second International Workshop on Issues of
  Sentiment Discovery and Opinion Mining}, page~10. ACM, 2013.

\end{thebibliography}

\end{document}